\documentclass[12pt,aps,prb]{revtex4}

\usepackage{amssymb,amsfonts,amsthm}

\usepackage{graphicx}
\usepackage{subfigure}

\usepackage{amssymb}
\usepackage{amsthm}
\usepackage{amsmath}
\usepackage{bm}             

\usepackage{psfrag}

\newcommand{\arsinh}{\mathop{\mathrm{arsinh}}} 
\DeclareMathOperator{\arcosh}{arcosh}

\theoremstyle{plain}

\theoremstyle{remark}

\newtheorem*{example}{Example}

\begin{document}

\title{Relativistic Aberration for Accelerating Observers}
\author{Robert Beig}
\email{Robert.Beig@univie.ac.at}
\author{J.\ Mark Heinzle}
\email{Mark.Heinzle@univie.ac.at}
\affiliation{Gravitational Physics, Faculty of Physics, University of Vienna,
A-1090 Vienna, Austria}

\date{\today}

\begin{abstract}

We investigate the effects of the 
aberration of light for a uniformly accelerating observer. 
The observer we consider 
is initially at rest with respect to a luminous
spherical object---a star, say---and then
starts to move away with constant acceleration.
The main results we derive are the following:
(i) The observer always sees an
initial increase of the apparent size of the object;
(ii) The apparent size of the object approaches a non-zero value
as the proper time of the observer goes to infinity.
(iii) There exists a critical value
of the acceleration such that
the apparent size of the object
is always increasing when the acceleration is super-critical.
We show that, while (i) is a purely non-relativistic effect, (ii) and (iii)
are effects of the relativistic aberration of light and
are intimately connected with the Lorentzian geometry
of Minkowksi spacetime.
Finally, the examples we present illustrate that, while more or less negligible in everyday life,
the three effects can be significant
in the context of space-flight.

\end{abstract}

\maketitle


\section{Introduction}

Imagine that after you have visited the Egyptian pyramids
you get in your car and leave. While you press the gas pedal,
in the rear view mirror you observe the pyramids
becoming smaller and smaller until finally they 
have shrunk to points.
(That this takes an infinite amount of time---and petrol---shall not bother us here.)
Interestingly enough, you are deceived.
What actually will happen is, first, that the pyramids in your
rear view mirror will grow in apparent size initially,
and second, that no matter how far you go, the apex angle
under which you observe these spectacular objects will never converge
to zero exactly
but remains positive even in the limit of infinite time.
In this paper we will show how these counterintuitive
facts arise from a combination of (non-relativistic and relativistic) 
aberration effects and
the assumed uniform acceleration of the observer.

The discovery and the correct 
description of the phenomenon of the aberration of light 
is one of the great moments in the history of physics and astronomy.
As early as 1725, James Bradley observed and measured 
the aberration of light emitted from ''fixed'' stars 
and correctly 
attributed it to the finite speed of light and the motion of the earth 
in its orbit around the sun.
(The first measurements resulted in the value $c\simeq 301 000 km/s$
for the speed of light.)
The relevant equations that describe the aberration of light 
on the level of
Newtonian/Galilean physics 
are good approximations as long as the 
involved velocities (of the observer and of the light source)
are much smaller than the speed of light.
However, it is clear that a 
theory that consistently describes the aberration of light must
take into account the invariance of the speed of light.
Such a theory is provided by special relativity.
In the context of relativistic physics, 
the 
aberration of light continues to be a
fascinating topic.
It attracts both the general public---intrigued, for instance, by
Mr.\ Tompkins' ride on his
relativistic bicycle\cite{Gamov}---
and the purely scientific readership, see, e.g., Ref.~\onlinecite{Penrose}.
Furthermore, the
visualization of relativistic aberration effects
has become an active field of research, which
poses challenges to both the theory
and its computer implementation, see~Ref.~\onlinecite{Wetal}
and references therein.

In this paper we consider aberration effects as 
seen by observers in a state of uniformly accelerated motion.
Uniform acceleration is
a topic discussed in every textbook
on special relativity\cite{Rindler};
in the Newtonian limit, we recover 
the uniformly accelerated motion of classical mechanics;
however, the relativistic generalization takes
into account the fact that there is an
(unattainable) limit to the observer's velocity: the speed of light.
Some aspects of the relativistic aberration of light for
accelerated observers have been discussed in, e.g.,
Refs.~\onlinecite{Scorgie} and~\onlinecite{LD}.
In this paper we analyze in detail the apparent
size of an object---a star, say---as seen
by a uniformly accelerated observer.
Our results add some unexpected and---to the
best of our knowledge---new features to
the phenomenon of the aberration of light.

The paper is structured as follows.
In Section~\ref{uniformaccel} we recapitulate the basic formulas
from relativistic kinematics which are necessary to describe
the motion of a uniformly accelerating observer.
In Section~\ref{relaberr} we give the formulas describing
the relativistic aberration effect
and apply these formulas to the situation under consideration.
The result is an implicit equation for the function $\theta(s)$,
which describes the apex angle $\theta$ under which
the observer sees the object at proper time $s$.
This equation can be analyzed rather easily in the limiting
cases of very small times (i.e., $s\rightarrow 0$) and
very large times (i.e., $s\rightarrow \infty$).
This analysis is performed in Sections~\ref{limitingcase1}
and~\ref{limitingcase2}.
We show that (i) $\theta(s)$ must always increase initially,
since its derivative at $s=0$ is always positive;
this fact is also briefly discussed in~Section V.B of Ref.~\onlinecite{MKA};
(ii) $\theta(s)$
always converges to a limit $\theta_\infty>0$ as $s\rightarrow \infty$.
In Section~\ref{asfctofs} we investigate in detail the
properties of the function $\theta(s)$. We show that there
exists a critical value of the acceleration such that,
in the case of under-critical accelerations,
$\theta(s)$ is increasing, assumes a maximum, and is decreasing again
until it reaches $\theta_\infty$ from above,
and, in the case of over-critical accelerations,
$\theta(s)$ is monotonically increasing until it
reaches $\theta_\infty$ from below.
Section~\ref{examples} contains a number of
examples. We apply the formulas we have derived
to concrete situations and find that the effects
we describe are negligible in everyday life, but
significant in the context of space-flight.
Finally, in Section~\ref{geom} we make contact with
more geometric considerations.
We show that our formulas are intimately connected with
geometric properties of the world lines of the observer
and the observed object in Minkowski spacetime.

\section{Uniformly accelerating observers}
\label{uniformaccel}

Let us begin by recapitulating some basic formulas
from relativistic kinematics\cite{Rindler}.
Let $(t,\vec{x})$ be an inertial system.
Consider an observer $\mathcal{O}$ that is at rest for $t\leqslant 0$ with
position $x= x_0>0$, $y=0$, $z= 0$.
At $t=0$ the observer begins to accelerate;
for $t > 0$, the observer then moves along the positive $x$-axis with
uniform (i.e., constant) linear acceleration $a>0$.
Since $a = \gamma^3 d v/d t$, where $\gamma= (1 - v^2)^{-1/2}$ is the usual
$\gamma$-factor, we obtain by simple integration that
\begin{subequations}\label{avxt}
\begin{alignat}{2}
\label{at}
a(t) & \equiv a = \mathrm{const} > 0\,, \\[0.5ex]
\label{vt}
v(t) & = \frac{a t}{\sqrt{1 + a^2 t^2}}\,, & \qquad  v(0) & = 0\,, \\[0.5ex]
\label{xt}
x(t) & = x_0 + \frac{1}{a} \left( \sqrt{1 + a^2 t^2} -1 \right)\,, & \qquad x(0) & = x_0 \,.
\end{alignat}
\end{subequations}
Note in this context that in order to simplify the formulas 
we choose units such that $c = 1$.
(Of course, for the applications in Section~\ref{examples} we use SI units.)
The proper time $s$ as measured by the observer's clocks is given by
\begin{equation}
s = \int_0^t \sqrt{1 - v^2(\tau)} \,d \tau = \frac{1}{a} \arsinh a t\:,
\quad \text{so that}\quad
t(s) = \frac{1}{a} \: \sinh a s \:.
\end{equation}
Expressed in proper time $s$, acceleration, velocity $v(s)$, and position $x(s)$ become
\begin{align}
\label{as}
a(s) & \equiv a = \mathrm{const} \,,
\tag{\ref{at}$^\prime$}\\[0.5ex]
\label{vs}
v(s) & = \tanh a s \,,
\tag{\ref{vt}$^\prime$}\\[0.5ex]
\label{xs}
x(s) & = x_0 + \frac{1}{a} \, \big( \cosh a s -1  \big) =
x_0 + \frac{2}{a}\: \sinh^2 \frac{a s}{2}\,.
\tag{\ref{xt}$^\prime$}
\end{align}
In four vector notation (where we suppress
the vanishing $y$- and $z$-components), we obtain
\begin{subequations}
\begin{alignat}{2}
\label{xmus}
x^\mu & = \frac{1}{a} \begin{pmatrix} \sinh a s \\ 
a x_0 + 2 \sinh^2 a s/2 \end{pmatrix} \\
\dot{x}^\mu & = \begin{pmatrix} \cosh a s \\ \sinh a s \end{pmatrix}
& & \dot{x}^\mu \dot{x}_{\mu} = 1 \\
\ddot{x}^\mu & = \begin{pmatrix} a \sinh a s \\ a \cosh a s \end{pmatrix}
& & \ddot{x}^\mu \ddot{x}_\mu = a^2
\end{alignat}
\end{subequations}
for the world line,
the four velocity and the four acceleration of the observer $\mathcal{O}$, respectively.
Note that we denote the derivative w.r.t.\ proper time by an overdot.
Finally, for later purposes we introduce an auxiliary function $\xi(s)$ which is given by
\begin{equation}\label{xis}
\xi(s) = \arcosh \frac{x(s)}{r}\,, \quad\text{so that}\quad
\dot{\xi} = \frac{1}{\sinh \xi} \frac{\dot{x}}{r}
= \frac{\dot{x}}{\sqrt{x^2-r^2}}\:.
\end{equation}
The function $\xi(s)$ is essentially a logarithmic measure for the distance $x(s)$.
Therefore, since $x(s)$ grows exponentially for large $s$, cf.~\eqref{xs}, $\xi(s)$
exhibits linear growth in the asymptotic regime $s\rightarrow \infty$,
in fact $\xi(s) = a s + \mathrm{l.o.t.}$ as $s\rightarrow \infty$;
this will be discussed in full detail below, see~\eqref{xxiasy}.

\section{Relativistic aberration}
\label{relaberr}

\begin{figure}[Ht]
\begin{center}
\subfigure[\,\,observer at rest]{\psfrag{r}[cc][cc][1][0]{$r$}
\psfrag{th}[cc][cc][1][0]{$\vartheta$}
\psfrag{x}[cc][cc][1][0]{$x$}
\includegraphics[width=0.4\textwidth]{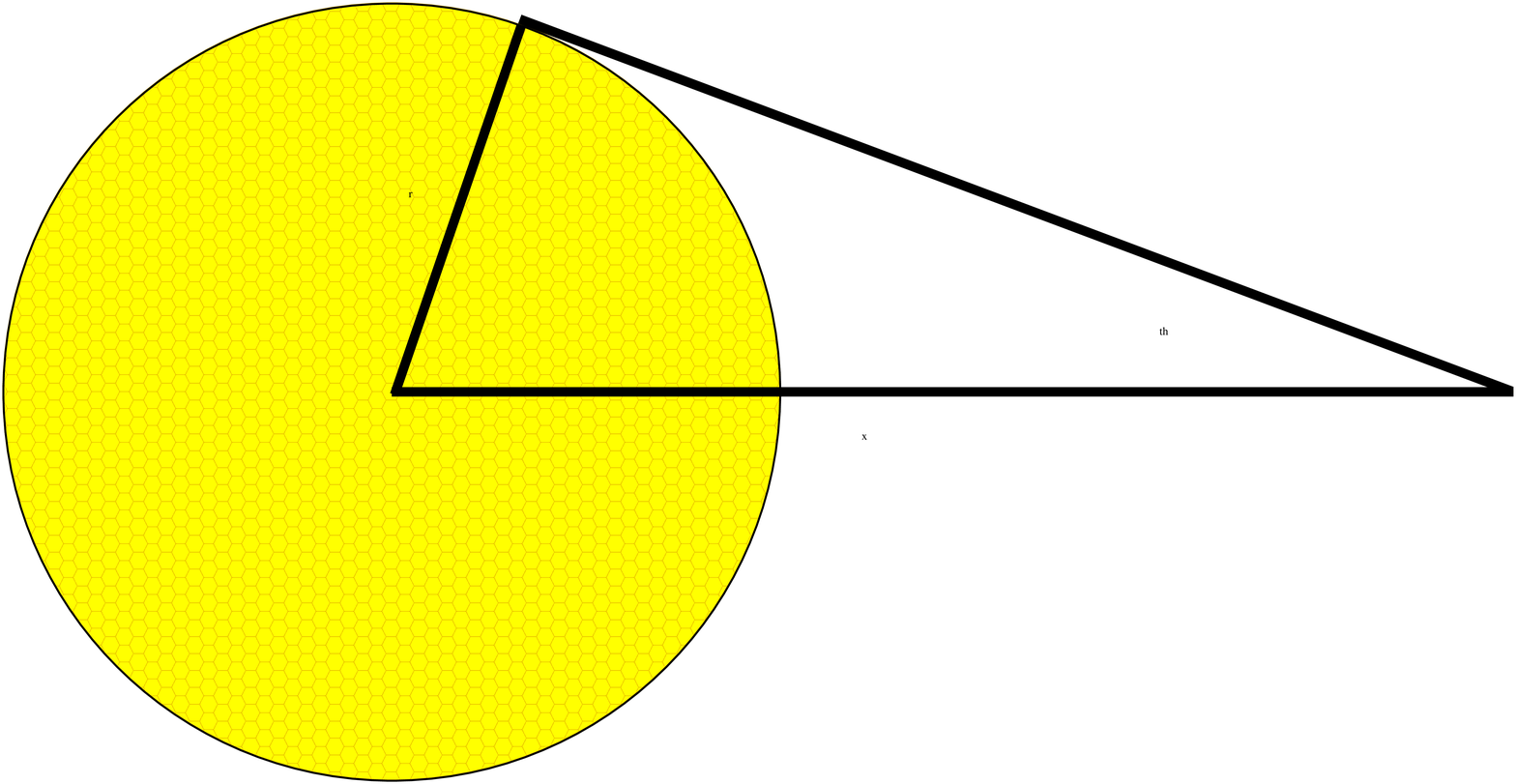}}
\qquad
\subfigure[\,\,moving observer]{\psfrag{r}[cc][cc][1][0]{$r$}
\psfrag{th}[cc][cc][1][0]{$\theta$}
\psfrag{x}[cc][cc][1][0]{$x$}
\includegraphics[width=0.4\textwidth]{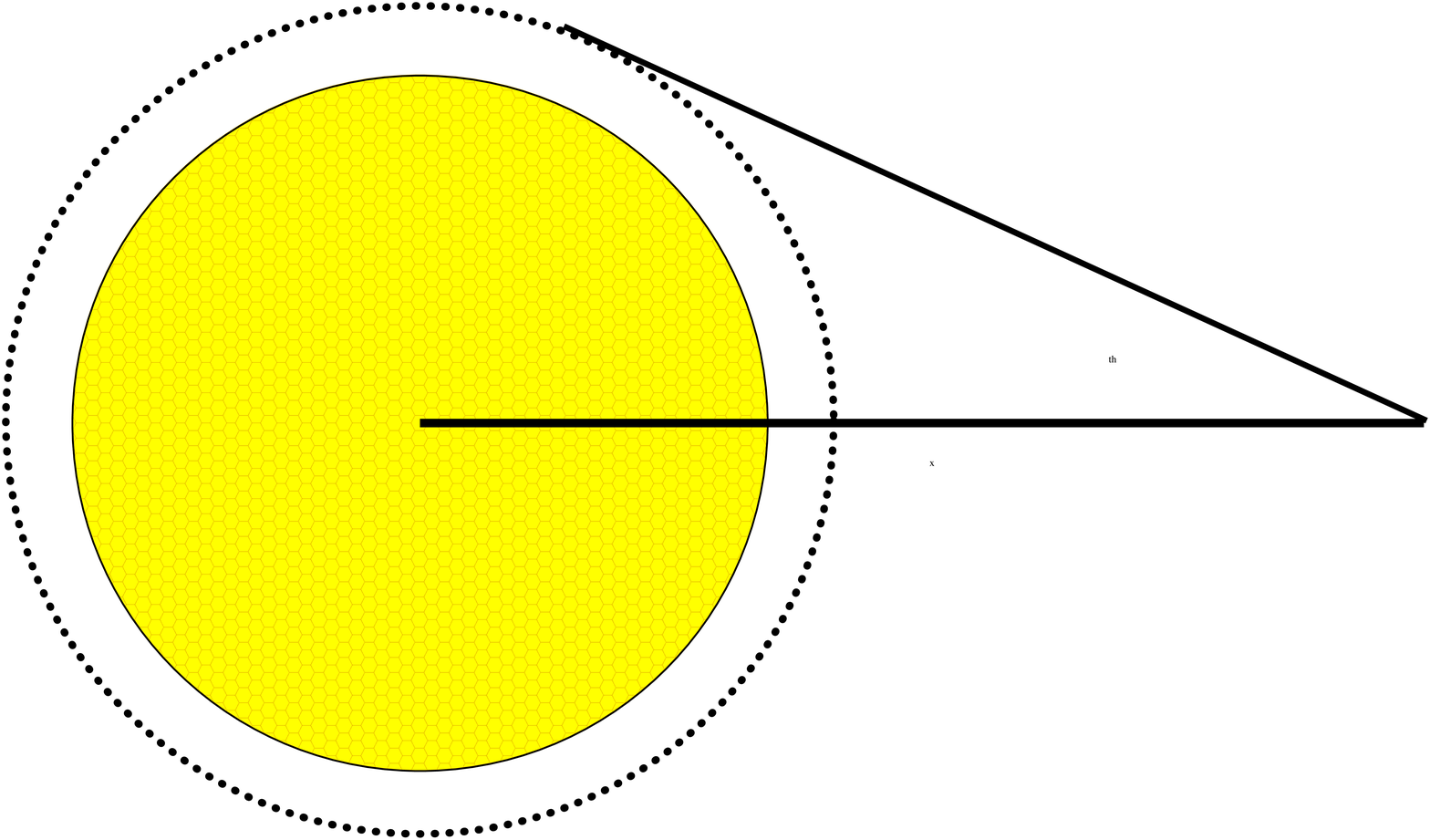}}
\caption{An observer at rest, whose distance to (the center of) the ball is $x$,
sees the ball at an angle $2 \vartheta = 2 \arcsin r/x$. For
a moving observer, aberration effects have to be taken into account;
the corrected apex angle is denoted by $\theta$.}
\label{picture}
\end{center}
\end{figure}

Consider now a ball of radius $r$, e.g., a star, that is located at $x=0$, $y=0$, $z=0$.
For an observer at a distance $x$ (from the center of the ball),
who is \textit{at rest} relative to the ball,
the ball appears at an apex angle $2 \vartheta$, where $\vartheta$ is given by
\begin{equation}\label{sinvartheta}
\sin \vartheta = \frac{r}{x}\:,
\end{equation}
see Figure~\ref{picture}.
Hence, for our observer $\mathcal{O}$,
for $s\leqslant 0$ ($t\leqslant 0$), the ball's apparent size
is $2 \vartheta_0 = 2 \arcsin r/x_0$.

Now, for an observer at the same distance $x$, whose relative velocity
w.r.t.\ the ball is non-vanishing, aberration effects must be taken
into account.
To be more specific, consider a moving observer
who moves at speed $v$ in
a direction away from the star, and
denote the apex angle at which the star appears for
this observer by $2 \theta$.
(In this context it is irrelevant whether the observer is in uniform
motion $v = \mathrm{const}$, or whether the velocity 
changes with time, i.e.,  $v = v(s)$.)
The special relativistic aberration formula applies and
$\theta$ is determined by either of the equivalent representations
of the relativistic aberration relation\cite{Rindler},
\begin{subequations}\label{ab}
\begin{align}
\label{ab1}
\cos \theta & = \frac{\cos\vartheta-v}{1- v \cos\vartheta} \:, \\[1ex]
\label{ab2}
\tan \frac{\theta}{2} & = \sqrt{\frac{1+v}{1-v}} \: \tan \frac{\vartheta}{2} \:, \text{ or} \\[1ex]
\label{ab3}
\sin \theta & = \sqrt{1-v^2}\:\frac{\sin \vartheta}{1- v \cos\vartheta} \:.
\end{align}
\end{subequations}
Note that equivalence between the formulas~\eqref{ab1} and~\eqref{ab2}
follows from the identity
$\tan \alpha = (1-\cos 2\alpha)/(\sin 2\alpha)$.
Since $\sin\theta$ is not bijective on $\theta\in[0,\pi)$,
to invert Equation~\eqref{ab3} it must be completed by
the additional requirement that
$\theta \in [0,\pi/2)$ if $\cos\vartheta > v$ and $\theta \in (\pi/2,\pi)$ if $\cos\vartheta < v$.
We will make us of these variants of the relativistic aberration formula in
different contexts.

For our accelerating observer $\mathcal{O}$
the distance $x = x(s)$ to the ball increases with proper time $s$;
simultaneously, the velocity $v = v(s)$ is increasing.
Accordingly,  the quantities on the r.h.\ side of~\eqref{ab} 
are functions of proper time $s$,
i.e., $\vartheta = \vartheta(s)$ and $v = v(s)$. 
The aberration formulas thus yield $\theta$ as a function $\theta(s)$.
(Note in this context that the very concept of the aberration of light 
depends solely on the four-velocities of two observers
and a null ray all at the same spacetime event, where, in our case, we have one 
(hypothetical) observer who is at rest w.r.t.\ the source
(and whose angle is $\vartheta$) and one 
observer $\mathcal{O}$ who recedes from the source (and
whose angle is $\theta$).
Therefore, in particular, only the (four-)velocity of $\mathcal{O}$ at
a particular event enters the formulas, so that it is irrelevant
whether $\mathcal{O}$ is accelerated or in uniform motion;
accordingly, the equations~\eqref{ab} 
can be applied to a particular instant of time $s$,
when the momentary velocity of $\mathcal{O}$ is $v = v(s)$.)

\begin{figure}[Ht]
\begin{center}
\psfrag{w}[cc][cc][1][0]{$w(\theta)$}
\psfrag{z}[lc][lc][1][0]{$\theta$}
\psfrag{a}[tc][cc][0.8][0]{$\frac{\pi}{2}$}
\psfrag{b}[tc][cc][0.8][0]{$\frac{3 \pi}{4}$}
\psfrag{c}[tc][cc][1][0]{$\pi$}
\psfrag{d}[bc][cc][0.8][0]{$\frac{\pi}{4}$}
\psfrag{2}[rc][rc][0.8][0]{$2$}
\psfrag{4}[rc][rc][0.8][0]{$4$}
\psfrag{-2}[rc][rc][0.8][0]{${-}2$}
\psfrag{-4}[rc][rc][0.8][0]{${-}4$}
\includegraphics[width=0.45\textwidth]{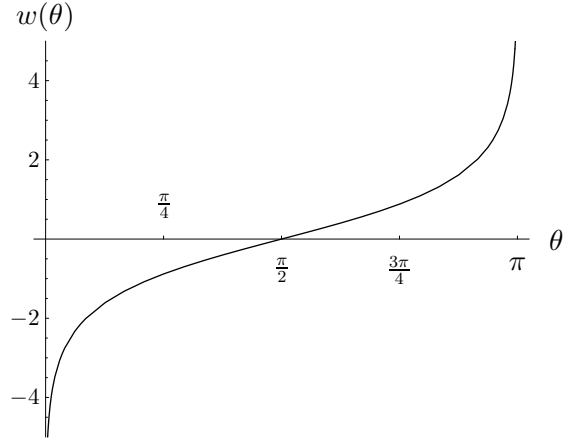}
\caption{A plot of the auxiliary function $w(\theta)$.}
\label{wofz}
\end{center}
\end{figure}

Since 
the distance $x = x(s)$ of the observer $\mathcal{O}$ to the ball increases,
the angle $\vartheta(s)$ decreases monotonically according to~\eqref{sinvartheta};
we may write~\eqref{sinvartheta} in the form
\begin{equation}\label{varthetaxi}
\tan \frac{\vartheta(s)}{2} = \frac{1}{r} \left( x(s) - \sqrt{x^2(s) -r^2}\right) =  e^{-\xi(s)}\,,
\end{equation}
where we have used~\eqref{xis}.
By definition, since $\mathcal{O}$ is moving w.r.t.\ to the ball for $s>0$ ($t>0$),
$\vartheta(s)$ is (half) the apex angle of the ball before taking into account
the aberration effect. The actual angle $\theta(s)$ at which $\mathcal{O}$
observes the ball is determined by~\eqref{ab2}. Since $v =\tanh a s$, see~\eqref{vs},
and thus $\sqrt{(1+v)/(1-v)} = e^{a s}$,
we find
\begin{equation}\label{thetaxi}
\tan \frac{\theta(s)}{2} = e^{a s - \xi(s)}\:.
\end{equation}
In order to simplify this equation we introduce an auxiliary function
\begin{subequations}
\begin{align}
& w : \alpha \in (0,\pi) \: \mapsto\: w(\alpha) \in \mathbb{R} \\
& w(\alpha) = \log \tan \frac{\alpha}{2}\:,\qquad
w^\prime(\alpha) = \frac{1}{\sin \alpha} \:,\qquad
w^{\prime\prime}(\alpha) = -\frac{1}{\sin \alpha \tan \alpha}\:,
\end{align}
\end{subequations}
see Figure~\ref{wofz}.
Using this function, Equation~\eqref{varthetaxi} becomes $w(\vartheta(s)) = -\xi(s)$ and~\eqref{thetaxi}
reads
\begin{equation}\label{wxis}
w\big(\theta(s)\big) = a s - \xi(s)\:.
\end{equation}
Since $\xi(s) \sim a s$
in the asymptotic regime $s\rightarrow \infty$, \eqref{wxis} becomes
particularly simple in this limit;
however,
we begin our analysis of the curve $\theta(s)$ 
by investigating the limiting case $s\rightarrow 0$ first.

\section{The Newtonian realm: the limiting case $\bm{s\rightarrow 0}$}
\label{limitingcase1}

The analysis of $\theta(s)$ in the limit of small $s$ can be performed in the context
of Newtonian/Galilean physics.
This is simply because, for small $s$, the velocity of the observer
is small compared to the speed of light, so that Newtonian physics
must hold (as a good approximation to relativistic physics).
In this section we study the limit of small $s$ first from a Newtonian perspective;
eventually, we return to the formalism developed in the previous sections.

Obviously, we have
\begin{equation*}
x(s) = \frac{a}{2} \:s^2
\end{equation*}
as the Newtonian limit of~\eqref{xs} and
the Newtonian aberration formula~Eq.~(3-110) of Ref.~\onlinecite{Gold}
\begin{equation*}
\cos \theta = \frac{\cos \vartheta - v}{\sqrt{1 - 2 v \cos \vartheta + v^2}}\:,
\end{equation*}
which coincides with the Newtonian limit $v\ll 1$ of the correct relativistic formula~\eqref{ab1} up
to second order in $v$,
\begin{equation}\label{aberapprox}
\cos \theta = \cos \vartheta - v \sin^2\vartheta  + O(v^2) \:.
\end{equation}
Recall that we have chosen
units so that $c=1$; hence $v$ corresponds to $v/c$ in 
conventional units.
In the formula~\eqref{aberapprox}, as before, $\vartheta$ is given by $\vartheta = \arcsin x/r$.

This Newtonian approximation leads to the correct qualitative understanding of
the behavior of $\theta(s)$ for small $s$.
The aberration formula~\eqref{aberapprox} contains the observer's velocity $v$ and the position $x$
(encoded in $\vartheta$). The velocity $v$ increases linearly with time
(for small times); however, the position $x$ increases merely quadratically.
Consequently, to first order, for small times, the change in position is negligible and 
does not enter our considerations. 
This leaves only the change in velocity and the associated
aberration effect that dominates the physics for small times:
We obtain
\begin{equation*}
\frac{d}{d s}\Big|_{s=0} \cos \theta = \left(-\sin \theta \,\big|_{s = 0}\right) \dot{\theta}\,\big|_{s=0} = 
- \dot{v}\big|_{s=0} \sin^2\vartheta\,\big|_{s = 0}\:,
\end{equation*}
which results in $\dot{\theta}\,\big|_{s=0} = a \sin\vartheta_0 > 0$.
Accordingly, the accelerating observer is confronted with an apex angle that is
increasing initially.
Note again that this aberration effect can be interpreted as 
a purely Newtonian effect that results from the particular geometric set-up of
the problem and the formulas of Newtonian mechanics.

The analysis is similar when we use the formalism developed in the previous sections.
We consider first the function $\xi(s)$; its derivatives evaluated at $s=0$ are given by
\begin{subequations}
\begin{alignat}{2}
\xi\,\big|_{s=0} & = \arcosh \frac{x}{r} \:\Big|_{s=0} & & =  \arcosh\frac{x_0}{r} \,,\\
\dot{\xi}\,\big|_{s=0} & = \frac{1}{\cos\vartheta} \:\frac{\dot{x}}{x} \:\Big|_{s=0} & &= 0\,, \\
\ddot{\xi}\,\big|_{s=0} & = \frac{1}{\cos\vartheta} \:\frac{\ddot{x}}{x} \:\Big|_{s=0} & &= \frac{1}{\cos\vartheta_0}\: \frac{a}{x_0}\,.
\end{alignat}
\end{subequations}
Hence, when we differentiate~\eqref{wxis} and use that $\theta|_{s=0} = \theta_0 = \vartheta_0$,
we obtain
\begin{equation}
w^\prime  \dot{\theta}\,\big|_{s=0} = a - \dot{\xi}\big|_{s=0} = a \:,
\end{equation}
which results in
\begin{equation}\label{dotthetapos}
\dot{\theta}\,\big|_{s=0} = a \sin\vartheta_0 > 0\:.
\end{equation}

We thus arrive at the following conclusion: Irrespective of the value of $a$, the angle
$\theta(s)$ increases with $s$ at least in some interval $s\in(0,*)$.
This is despite the fact that the distance between the observer $\mathcal{O}$
and the ball increases with $s$ (so that $\vartheta(s)$ is obviously decreasing).
This result comes as a surprise at first. Everyday experience
suggests that when I press the gas pedal in my car while observing an object in
my rear view mirror, nothing spectacular happens;
on the contrary, in my experience the object's size is constantly decreasing.
Equation~\eqref{dotthetapos} shows that this is not true; however,
in our examples below we will see that the interval
$(0,*)$, in which $\theta(s)$ is increasing, spans merely a tiny fraction of a second,
and moreover, the increase in $\theta$ is only marginal;
these facts remove the apparent ``contradiction'' between~\eqref{dotthetapos} and
everyday experience.

To complete the analysis of $\theta(s)$ for
small $s$, we
compute the second derivatives of~\eqref{wxis}.  We obtain
\begin{equation}
w^{\prime\prime} \dot{\theta}^2 + w^\prime \ddot{\theta}\,\big|_{s=0} = -\ddot{\xi}\,\big|_{s=0}\:,
\end{equation}
and we infer that
\begin{equation}
\ddot{\theta}\big|_{s=0} = -\frac{a}{x_0} \tan\vartheta_0 \left( 1 - a x_0 \cos^2 \vartheta_0\right)\:.
\end{equation}
The sign of $\ddot{\theta}|_{s=0}$ depends on the value of $a$ (where we consider
$r$ and $x_0$ as given).
For small values of $a$, $\ddot{\theta}\big|_{s=0}<0$.
This is an indication of the above-mentioned fact that
positivity of $\dot{\theta}$ is only short-lived.
If $a$ is sufficiently large, however, then $\ddot{\theta}|_{s=0}>0$.
This suggests that, for large $a$, $\dot{\theta}(s)$ might remain positive (and thus $\theta(s)$ increasing)
for longer
times (or even forever).
Before we analyze these issues in detail we consider the second limiting case of~\eqref{wxis},
the case $s\rightarrow \infty$.

\section{A relativistic effect: the limiting case $\bm{s\rightarrow \infty}$}
\label{limitingcase2}

Equation~\eqref{xs} implies that
$x(s)$ behaves asymptotically according to
\begin{subequations}\label{xxiasy}
\begin{equation}
x(s) = \frac{e^{a s}}{2a} + \left( x_0 - \frac{1}{a}\right) + \frac{e^{-a s}}{2 a} =
\frac{e^{a s}}{2 a} \left( 1+ O(e^{-a s}) \right)
\end{equation}
so that $\xi(s)$, see~\eqref{xis}, becomes
\begin{equation}\label{xiasy}
\xi(s) = \arcosh \frac{x(s)}{r} = \log \frac{2 x(s)}{r} -\frac{1}{4} \frac{r^2}{x^2} + O\Big( \big(\frac{r}{x}\big)^4 \Big)
= a s - \log a r +O(e^{-a s})
\end{equation}
\end{subequations}
in the asymptotic regime $s\rightarrow \infty$.
This equation describes the linear growth of $\xi(s)$ in the asymptotic regime
alluded to before.
Clearly, the equation $w\big(\vartheta(s)\big) = -\xi(s)$ implies that
\begin{equation}
\vartheta(s) \xrightarrow{s\rightarrow \infty} \vartheta_\infty = 0 \:.
\end{equation}
This is not true, however, for the actual apex angle $\theta(s)$:
Equation~\eqref{wxis} becomes
\begin{subequations}
\begin{align}
w\big(\theta(s)\big) & =  a s - \xi(s) = \log a r + O(e^{-a s})\,,
\intertext{which in turn entails}
\tan \frac{\theta(s)}{2} & = a r  + O(e^{-a s})\qquad (s\rightarrow \infty)\:.
\end{align}
\end{subequations}
We conclude that in the limit $s\rightarrow \infty$, $\theta(s)$ converges
to a limiting value which we denote by $\theta_\infty$:
\begin{equation}\label{thetainf}
\theta(s) \xrightarrow{s\rightarrow \infty} \theta_{\infty} = 2 \arctan a r = \arccos \frac{1 - a^2 r^2}{1 + a^2 r^2} \:.
\end{equation}
This is a somewhat counterintuitive conclusion. Despite the fact that
the observer $\mathcal{O}$ is at infinite distance,
the object appears at the finite apex angle $2 \theta_\infty$.
For future purposes we note that $\theta_\infty$ can also be represented by the
relation
\begin{equation}
\sin \theta_{\infty} = \frac{2 a r}{1+ a^2 r^2}
\tag{\ref{thetainf}${}^\prime$}
\end{equation}
with the additional prescription that $\theta_\infty \in (0,\pi/2]$ if $a r \leq 1$,
and $\theta_\infty \in [\pi/2,\pi)$
if $a r \geq 1$.

To avoid misconceptions, we remark in passing that although the apparent size of the object reaches a constant
in the limit $s\rightarrow \infty$, clearly, the object becomes infinitely faint,
since the luminosity decreases to zero.

\section{The apex angle $\bm{\theta(s)}$ as a function of proper time $\bm{s}$}
\label{asfctofs}

In the following we analyze in detail the curve $\theta(s)$ given by~\eqref{wxis};
in particular we investigate whether this function has extremal points.
Differentiating~\eqref{wxis} we obtain
\begin{equation}
w^\prime(\theta) \,\dot{\theta} = \frac{1}{\sin\theta}\, \dot{\theta} = a -\dot{\xi} = a  - \frac{\dot{x}}{\sqrt{x^2-r^2}}\:.
\end{equation}
Since
\begin{equation}
\dot{x} = \sinh a s = 2 \sinh \frac{a s}{2} \:\sqrt{1 + \sinh^2 \frac{a s}{2}}
\end{equation}
we infer
\begin{equation}
\dot{\theta} = \sin\theta \left[ a -
\frac{2 \sinh \frac{a s}{2}\:\sqrt{1 + \sinh^2 \frac{a s}{2}}}{ \sqrt{\left(x_0+\frac{2}{a} \sinh^2\frac{a s}{2}\right)^2 - r^2}}\right]\:.
\end{equation}
By simple algebraic manipulations we subsequently find
that
\begin{equation}\label{curvdiss}
\dot{\theta}(s) = 0 \qquad \text{if and only if} \qquad
\sinh \frac{a s}{2}  = 
\frac{a}{2}\,\frac{\sqrt{x_0^2-r^2}}{\sqrt{1- a x_0}} \:.
\end{equation}
A solution $s$ of this equation exists under the condition
that $\sqrt{1-a x_0}$ be real and non-zero, i.e., if and only if
$a x_0 < 1$.
Hence, assuming $a x_0 <1$, we see that
$\dot{\theta}(s)$ has a unique zero, so that
$\theta(s)$ possesses a unique extremum.
This extremum is evidently
a maximum because $\dot{\theta}|_{s=0} > 0$.
In contrast, when $a$ is sufficiently large, i.e., when $a x_0 \geqslant 1$,
then $\theta(s)$ does not possess any extrema;
since it is monotonically increasing
for small $s$, it must then be a monotonically increasing function
for all $s \in (0,\infty)$.
The maximum of $\theta(s)$ is thus attained at infinity, where
$\theta = \theta_\infty$.

Consider the case $a x_0 <1$ and
let us use a subscript $(*)_m$ to denote the values of the variables at the maximum of the apex angle.
Equation~\eqref{curvdiss} translates to
\begin{subequations}
\begin{equation}\label{sm}
s_m = \frac{2}{a} \arsinh \frac{a}{2}\,\frac{\sqrt{x_0^2-r^2}}{\sqrt{1- a x_0}} \:.
\end{equation}
This can be used in~\eqref{vs} and~\eqref{xs} to obtain
\begin{equation}
x_m = x_0 + \frac{a}{2} \, \frac{x_0^2-r^2}{1- a x_0} \:, \qquad
v_m = \frac{2 \sqrt{1 + \sinh^2 \frac{a s_m}{2}} \:\sinh \frac{a s_m}{2}}{1 + 2 \sinh^2 \frac{a s_m}{2}}\:.
\end{equation}
To obtain $\theta_m$ it is simplest to use $\sin \vartheta_m = r/x_m$ and
compute $\theta_m$ in the representation~\eqref{ab3}.
This yields
\begin{equation}\label{sinthetam}
\sin \theta_m = \frac{\sqrt{1-v_m^2}\:\sin\vartheta_m}{1 -v_m \cos \vartheta_m}
= \frac{r}{x_0 - \frac{a}{2} (x_0^2 -r^2)}\:.
\end{equation}
\end{subequations}
The denominator is not manifestly positive; however,
our assumption $a x_0 <1$ guarantees positivity, since
\[
x_0 - \frac{a}{2} (x_0^2 -r^2) > 0 \quad\Leftrightarrow\quad
a x_0 ( 1 -a x_0) > -\frac{a^2}{2}\, (r^2 + x_0^2)\:,
\]
which is true by $a x_0 <1$.
For small $a$, the r.h.s.\ of~\eqref{sinthetam} is clearly less than one,
and the angle $\theta_m$ takes a value in $(0,\pi/2)$.
A straightforward manipulation shows that this remains true for all $a$ such that $a x_0 <1$;
indeed,
\[
\frac{r}{x_0 - \frac{a}{2} (x_0^2 -r^2)} <1 \quad \Leftrightarrow \quad
a r \left(1 -\frac{a r}{2}\right) < a x_0 \left( 1 -\frac{a x_0}{2}\right)\:,
\]
which is in turn equivalent to $r< x_0$ and thus holds trivially.
It follows that the inversion of~\eqref{sinthetam} always yields a value
for the angle $\theta_m$ that is less than $\pi/2$.

Let us collect the results
on the behavior of the apex angle $\theta(s)$ at which
the observer $\mathcal{O}$ sees the star.
Initially, at time $s=0$, we have $\sin \theta_0 = r/x_0$;
with increasing $s\in (0,*)$ the apex angle $\theta(s)$ increases.
If $a x_0 < 1$ then $\theta(s)$ is increasing for $s\in (0,s_m)$,
where $s_m$ is given by~\eqref{sm}, reaches a maximum $\theta(s_m) = \theta_m$,
and decreases again for $s\in (s_m,\infty)$.
If $a x_0 \geqslant 1$ then
$\theta(s)$ is monotonically increasing for all $s\in (0,\infty)$.
In any case, $\theta(s)$ converges to a limit $\theta_\infty$ as $s \rightarrow \infty$.
The corresponding values of the apex angle are given by
\begin{subequations}\label{thetaminf1}
\begin{alignat}{2}
\sin \theta_m & = \frac{r}{x_0 - \frac{a}{2} (x_0^2 -r^2)} \qquad\quad & & ( a x_0 < 1)  \\[1ex]
\label{sininf}
\sin \theta_\infty & = \frac{2 a r}{1 + (a r)^2}\:, & & 
\end{alignat}
\end{subequations}
where the latter equation must be supplemented by the prescription
that $\theta_\infty \in (0,\pi/2)$, if $a r < 1$, and
$\theta_\infty \in [\pi/2,\pi)$, if $a r \geqslant 1$.

\begin{figure}[Ht]
\begin{center}
\psfrag{s}[cc][cc][1][0]{$s$}
\psfrag{t}[cc][cc][1][0]{$\theta(s)$}
\psfrag{2}[rc][rc][0.8][0]{$2^\circ$}
\psfrag{3}[rc][rc][0.8][0]{$3^\circ$}
\psfrag{4}[rc][rc][0.8][0]{$4^\circ$}
\psfrag{5}[rc][rc][0.8][0]{$5^\circ$}
\includegraphics[width=0.6\textwidth]{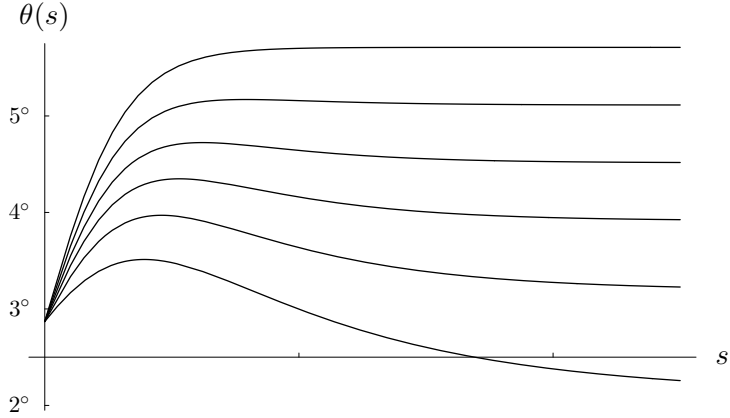}
\caption{The apex angle function $\theta(s)$ for a series of under-critical values of $a$.
Since $a x_0 < 1$, the curve $\theta(s)$ is increasing for $s\in (0,s_m)$,
reaches a maximum $\theta(s_m) = \theta_m$,
and decreases again for $s\in (s_m,\infty)$. The topmost curve corresponds to the critical value $a x_0 =1$.
If $a x_0 \geq 1$, $\theta(s)$ is monotonically increasing.}
\label{varyinga}
\end{center}
\end{figure}

\begin{figure}[Ht]
\begin{center}
\psfrag{a}[lc][lc][1][0]{$a x_0$}
\psfrag{t}[cc][cc][1][0]{$\theta_m/\theta_\infty$}
\psfrag{tm}[cc][cc][1][0]{$\theta_m$}
\psfrag{ti}[cc][cc][1][0]{$\theta_\infty$}
\psfrag{0.5}[tc][tc][0.8][0]{$0.5$}
\psfrag{1}[tc][tc][0.8][0]{$1$}
\includegraphics[width=0.45\textwidth]{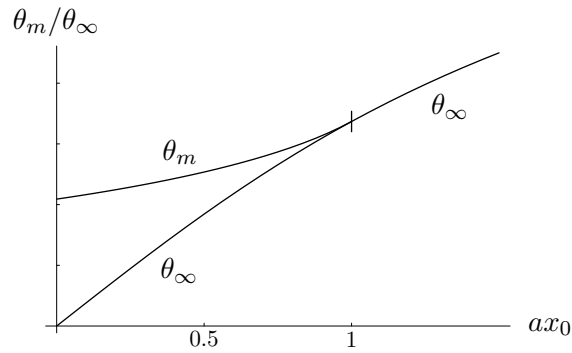}
\caption{When $r$ and $x_0$ are considered as fixed, $\theta_m$ and $\theta_\infty$
can be regarded as functions
of the acceleration $a$, cf.~\eqref{thetaminf1}. For small values of $a$, i.e., for $a x_0 <1$,
since the curve $\theta(s)$ attains the maximum value $\theta_m$ at $s = s_m < \infty$,
$\theta_m$ is larger than $\theta_\infty$. For $a x_0 \geqslant 1$, the curve $\theta(s)$
is increasing for all $s$, hence $\theta_m$ is undefined and $\theta_\infty$ represents the maximum.}
\label{thetafig}
\end{center}
\end{figure}

Figure~\ref{varyinga} shows a plot of $\theta(s)$ for a series of under-critical values
of $a$ (i.e., $a x_0 <1$) and the critical value $a x_0 = 1$.
For a plot of $\theta_m$ and $\theta_\infty$, again for given $r$, $x_0$,
in dependence of $a$ (where $a$ is measured in units of $a x_0$),
see Figure~\ref{thetafig}.

A simple consequence of these considerations is that
\begin{subequations}
\begin{alignat}{2}
& \sin \theta_m = \sin \theta_\infty & \quad & \Leftrightarrow \quad a x_0 = 1\,;
\intertext{this can also be checked by a direct calculation.
Furthermore,}
& \sin \theta_0 = \sin \theta_\infty & \quad & \Leftrightarrow \quad a r =
\tan\frac{\vartheta_0}{2} = \frac{x_0}{r} - \sqrt{\frac{x_0^2}{r^2}-1}\,;
\end{alignat}
\end{subequations}
in this case, clearly $a x_0 < 1$ holds.

\section{Applications and examples}
\label{examples}

In order to discuss some particular examples and applications
we now introduce units.
Equations~\eqref{vs} and~\eqref{xs} become
\begin{equation}
v(s) = c \tanh \frac{a s}{c} \qquad \text{and} \qquad
x(s) = x_0 + \frac{2 c^2}{a}\: \sinh^2 \frac{a s}{2 c}\:.
\end{equation}
The curve $\theta(s)$ has a maximum when $a x_0 < c^2$;
this maximum occurs at proper time
\begin{equation}\label{smunits}
s_m = \frac{2 c}{a} \arsinh \frac{a}{2 c}\,\frac{\sqrt{x_0^2-r^2}}{\sqrt{c^2- a x_0}} \:.
\end{equation}
If $a x_0 \geqslant c^2$, then $\theta(s)$ is monotonically increasing for all $s$.
The key characteristics of the curve $\theta(s)$ are the following:
\begin{subequations}\label{thetaunits}
\begin{alignat}{3}
& \sin \theta_0 && = \frac{r}{x_0} \\[1ex]
& \sin \theta_m && = \frac{r}{x_0 - \frac{a}{2 c^2} (x_0^2 -r^2)}  \qquad\quad & & ( a x_0 < c^2)\\[1ex]
& \sin \theta_\infty && = \frac{2 a r c^2}{(a r)^2+c^4}\, & & 
\end{alignat}
\end{subequations}
where $\theta_\infty \in (0,\pi/2)$ if $a r < c^2$ and $\theta_\infty \in [\pi/2,\pi)$ else.

\begin{example}
The acceleration of a formula one racing car is of the order of magnitude of
$1 g$ (where $g \simeq 10 m/s^2$). Let us consider a formula one
driver who accelerates uniformly on a straight track, while
he observes a distant mountain in his rear view mirror.
We assume the mountain's height to be $r = 3\, km$ and
the initial distance to be $x_0 = 10 \,km$;
accordingly, $\theta_0 \simeq 17.5^\circ$.
Using~\eqref{smunits} and~\eqref{thetaunits} yields the following:
$s_m \simeq 3.2\cdot 10^{-5} s$, $\theta_m \simeq \theta_0 +3.3 \cdot 10^{-8}\:{}^{\prime\prime}$,
$\theta_\infty \simeq 1.4\cdot 10^{-7}\:{}^{\prime\prime}$.
It is thus evident that the driver is unaware of any aberration effect.
\end{example}

\begin{figure}[Ht]
\begin{center}
\subfigure{%
\psfrag{s}[cc][cc][1][0]{$s$}
\psfrag{t}[cc][cc][1][0]{$\theta$}
\psfrag{500}[tc][tc][0.8][0]{$500s$}
\psfrag{1000}[tc][tc][0.8][0]{$1000s$}
\psfrag{1500}[tc][tc][0.8][0]{$1500s$}
\psfrag{31.99}[rc][rc][0.8][0]{$31.99^\prime$}
\psfrag{31.995}[rc][rc][0.8][0]{$31.995^\prime$}
\psfrag{32}[rc][rc][0.8][0]{$32^\prime$}
\includegraphics[width=0.4\textwidth]{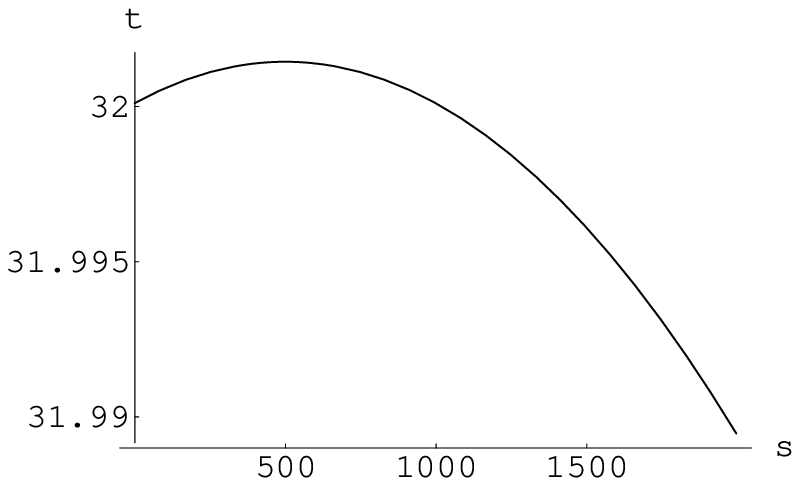}}
\qquad
\subfigure{%
\psfrag{s}[cc][cc][1][0]{$s$}
\psfrag{t}[cc][cc][1][0]{$\log\theta/\theta_{\infty}$}
\psfrag{5000000}[tc][tc][0.8][0]{$5\cdot 10^6s$}
\psfrag{10000000}[tc][tc][0.8][0]{$1\cdot 10^7s$}
\psfrag{2}[rc][rc][0.8][0]{$2$}
\psfrag{4}[rc][rc][0.8][0]{$4$}
\psfrag{6}[rc][rc][0.8][0]{$6$}
\includegraphics[width=0.4\textwidth]{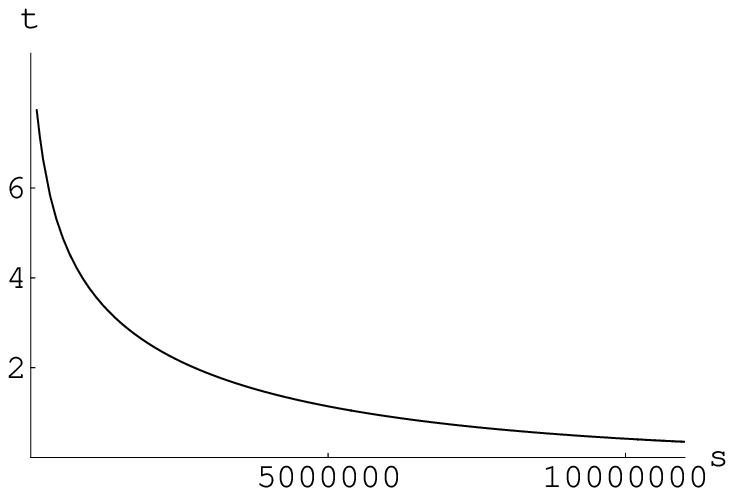}}
\caption{The function $\theta(s)$ for an observer accelerating
away from the sun, where $x_0$ is one astronomical unit.
We have chosen $a = 5 g$.}
\label{numbers}
\end{center}
\end{figure}

\begin{example}
The radius of the sun is approximately $r = 7 \cdot 10^8 m$.
Consider a rocket starting from earth or the moon,
which uniformly accelerates away from the sun.
In this example, $x_0$ is one astronomical unit, i.e., $x_0 \simeq 1.5\cdot 10^{11} m$.
Suppose that $a= 5 g$, which
rockets can achieve today (and which is tolerable by humans---at least for short times).
Then
\begin{equation}
2 \theta_0 \simeq 32^\prime \,,
\qquad
2 \theta_m \simeq 2 \theta_0 +  0.08^{\prime\prime}\,,
\qquad
s_m \simeq 500s\,,
\qquad
\theta_{\infty} \simeq 0.32^{\prime\prime}\,;
\end{equation}
see Figure~\ref{numbers}.
The effect is thus not particularly big but clearly measurable.
\end{example}

Before we discuss another example, let us consider the equations in the limit
of small accelerations $a$.
Equation~\eqref{smunits} becomes
\begin{equation}\label{smunits2}
s_m = c^{-1} \sqrt{x_0^2 -r^2} \:\, \Big[1 + O(a x_0/c^2)\Big]\:,
\end{equation}
and~\eqref{thetaunits} results in
\begin{subequations}\label{thetaunits2}
\begin{align}
2 \theta_m - 2\theta_0 & = \frac{r a}{c^2}\:\sqrt{1-\frac{r^2}{x_0^2}} \:\, \Big[1 + O(a x_0/c^2)\Big] \\
2 \theta_\infty & = \frac{4 r a}{c^2}\:  \big[1 + O(a^2 r^2/c^4)\big]
\end{align}
\end{subequations}
If $x_0$ is larger than $r$ by an order of $10$, say, then $\sqrt{1-r^2/x_0^2} \simeq 1$,
and~\eqref{smunits2} and~\eqref{thetaunits2} become
\begin{equation}
s_m \simeq \frac{x_0}{c}\:,
\qquad\qquad
2 \theta_m -2 \theta_0 \simeq \frac{r a}{c^2} \simeq \frac{1}{4} \:(2\theta_\infty)\:.
\end{equation}
These formulas confirm that aberration effects are negligible in
everyday life.
For space-travelers in the not so far future, however, aberration
will be ubiquitous, as will be seen in following example.

\begin{figure}[Ht]
\begin{center}
\psfrag{s}[cc][cc][1][0]{$s$}
\psfrag{t}[cc][cc][1][0]{$2\theta$}
\psfrag{500000000}[tc][tc][0.8][0]{$5\cdot 10^8s$}
\psfrag{1000000000}[tc][tc][0.8][0]{$10^9s$}
\psfrag{8}[rc][rc][0.8][0]{$8^\circ$}
\psfrag{12}[rc][rc][0.8][0]{$12^\circ$}
\psfrag{16}[rc][rc][0.8][0]{$16^\circ$}
\psfrag{20}[rc][rc][0.8][0]{$20^\circ$}
\psfrag{24}[rc][rc][0.8][0]{$24^\circ$}
\includegraphics[width=0.6\textwidth]{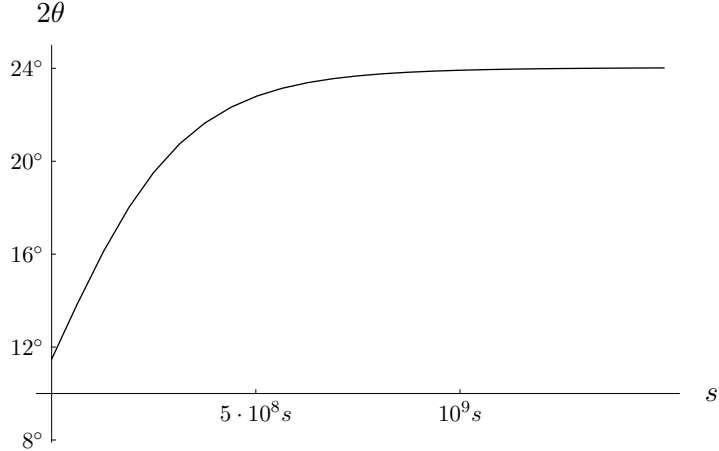}
\caption{The function $\theta(s)$ for an accelerating observer that
is initially at ten light years distance from an object with a radius of one light year.
We have chosen a moderate acceleration of $a = 1 g$.}
\label{numbers2}
\end{center}
\end{figure}

\begin{example}
Let $r$ be one light year, i.e., $r \simeq 9.5 \cdot 10^{15} m$.
This is a typical radius of a (small) SNR (supernova remnant).
Consider a spaceship at a distance of $x_0 = 10 r$ to the object
and suppose that the engines of the spaceship provide
a steady acceleration of $a= 1 g$. (This is a good choice to make
the spaceship an earth-like place for the space-farers.)
The following data for $\theta(s)$ emerges:
Since $a x_0/c^2 > 1$, the curve is monotonically increasing for all $s$
from $\theta_0$ to $\theta_\infty$, where
\begin{equation}
2 \theta_0 \simeq 11.5^{\circ} \,\qquad\qquad
2 \theta_{\infty} \simeq 24.0^{\circ}\:;
\end{equation}
see Figure~\ref{numbers2}.
\end{example}

\section{Aberration, acceleration, and Minkowski geometry}
\label{geom}

We conclude this paper by discussing the connections between
the results we obtained on the properties of $\theta(s)$
and the geometry in Minkowski spacetime of the
problem we consider. To simplify the equations we again
use units such that $c=1$.

\begin{figure}[Ht]
\begin{center}
\subfigure[\,\,$a x_0 <1$]{%
\psfrag{star}[tc][tc][1][0]{star}
\psfrag{obs}[tc][tc][1][0]{obs.}
\psfrag{t}[rc][rc][0.8][0]{$t_\infty$}
\label{rh1}\includegraphics[width=0.27\textwidth]{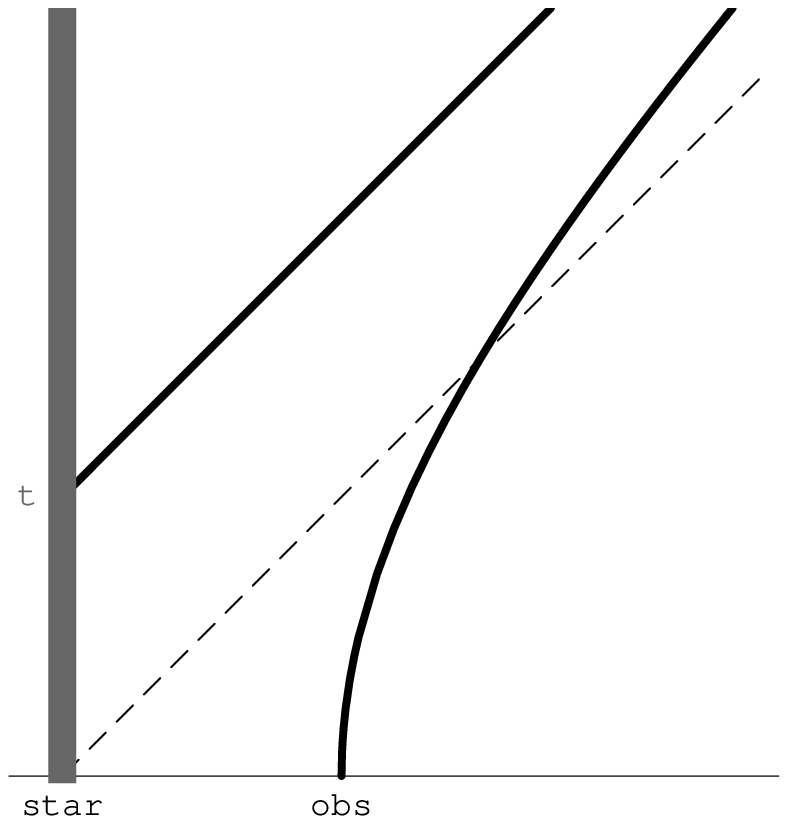}}
\qquad
\subfigure[\,\,$a x_0 = 1$]{%
\psfrag{star}[tc][tc][1][0]{star}
\psfrag{obs}[tc][tc][1][0]{obs.}
\label{rh2}\includegraphics[width=0.27\textwidth]{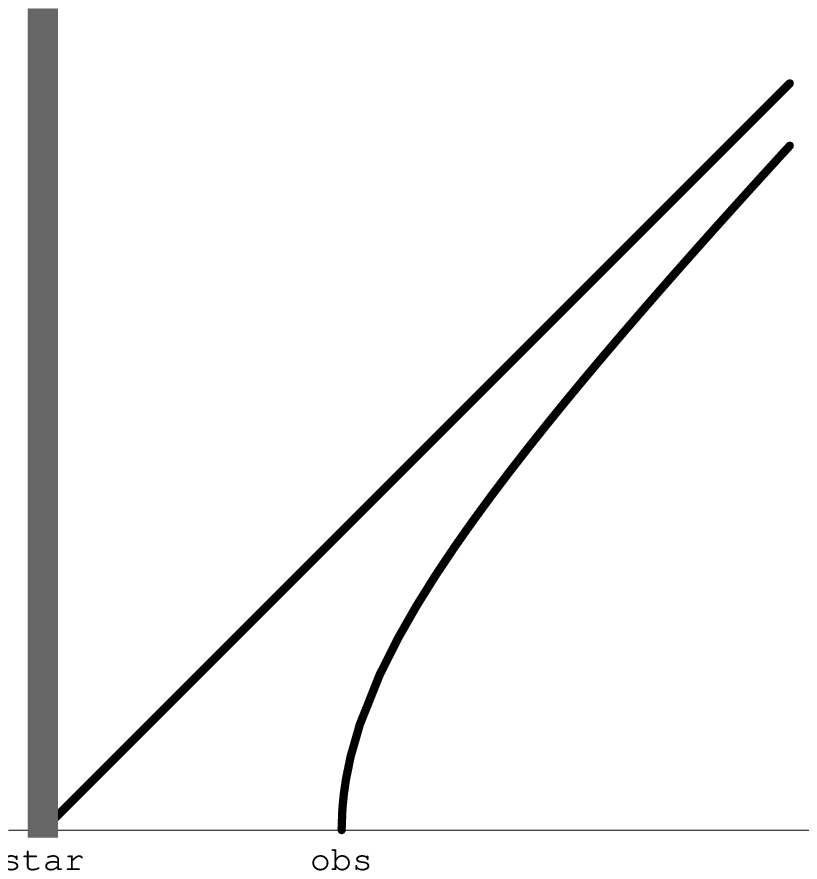}}
\qquad
\subfigure[\,\,$a x_0 > 1$]{%
\psfrag{star}[tc][tc][1][0]{star}
\psfrag{obs}[tc][tc][1][0]{obs.}
\label{rh3}\includegraphics[width=0.27\textwidth]{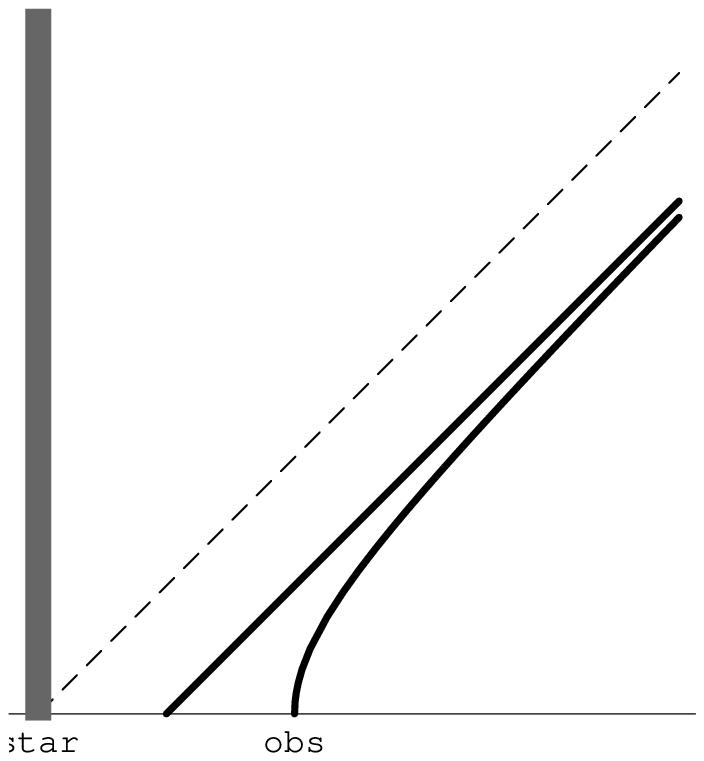}}
\caption{In this Minkowski diagram the star is represented by a straight world line/tube at
$\vec{x} = 0$ (gray line), while the uniformly accelerating observer is described by
an equilateral hyperbola (black solid curve);
the asymptote of this hyperbola is a null line (black solid line).
The dotted line is the Rindler horizon of the event $(t=0,\vec{x} = 0)$.}
\label{rhi}
\end{center}
\end{figure}

In Minkowski spacetime
the star is given by the world tube $\{(t,\vec{x}) \:|\: t\in\mathbb{R},\|\vec{x}\| \leqslant r\}$;
the center of the star is simply $\vec{x} = 0$.
On the other hand, the observer's world line
(for $s\geqslant 0$) is the upper branch of
the equilateral hyperbola
\begin{equation}\label{hyp}
\Big[ x - x_0 +  \frac{1}{a} \Big]^2 - t^2 = \frac{1}{a^2}\:,
\end{equation}
which is a simple consequence of~\eqref{xmus}. We choose to write~\eqref{hyp}
in the form
\begin{equation}
\Big[ x + \frac{1}{a} (1-a x_0)\Big]^2 - t^2 = \frac{1}{a^2}\:.
\tag{\ref{hyp}${}^\prime$}
\end{equation}
Accordingly, the hyperbola's asymptote is given by the straight line
\begin{equation}\label{nullasy}
\Big[ x + \frac{1}{a} (1-a x_0)\Big] - t = 0\:,
\end{equation}
which is a \textit{null line}.
The existence of this null asymptote
is particularly significant: it means that
the observer approaches null infinity
(as opposed to timelike infinity)
as $s\rightarrow \infty$.
Therefore, the causal past of the accelerating observer
(the boundary of which is given by the null asymptote)
does not comprise the entire Minkowski spacetime, but only a part of
it. For instance, the null asymptote intersects
the world line $\vec{x} = 0$ of the
(center of the) star at
\begin{equation}\label{tinf}
t = t_\infty = \frac{1}{a} \: \big(1 - a x_0\big) \:;
\end{equation}
see Figure~\ref{rhi}.
This time defines the limit up to which
the accelerating observer can see the star.
Light emitted from the star later than $t_\infty$ cannot
reach the observer, since it is outside of his causal past.
(In this context we have suppressed the fact that the star
is a finitely extended body. Taking this into account
one finds that $t_\infty$ depends on the part of
the surface of the star from which the light is emitted.
The part of the surface closest to the observer is
given by the world line $\{(t,r,0,0)\,|\,t\in\mathbb{R}\}$;
instead of~\eqref{tinf} one obtains $t_\infty^\prime = r + t_\infty$.
For light emitted from the part of the surface
that the observer sees as the star's contour, however,
\eqref{tinf} is correct.
Note that these subtleties disappear when we consider
a simpler geometry, namely a luminous rod instead
of a luminous ball.)
In the case of small accelerations, i.e.,  for $a x_0 < 1$,
the limiting time $t_\infty$ is positive;
for large accelerations, i.e., when $a x_0 > 1$,
$t_\infty$ is negative.
The case $a x_0 = 1$ is special: $t_\infty = 0$.
In this case, the light ray emitted by the star at $t = 0$
coincides with the asymptote of the observer's world line,
so that the observer can keep track of the star's
destiny up to $t =0$, which is exactly the time when
the observer began to accelerate.

Applying the concept of the \textit{Rindler horizon}
to our problem reveals further interesting aspects
and connections between our results and the underlying
spacetime geometry.
The Rindler horizon of an event is defined as
the future null cone of that event;
the chronological future of the event (which is the interior
of the null cone) is said to lie within the Rindler horizon.
Communication between the inside and the outside
is only one-way, since events within the Rindler horizon
cannot have any influence on events outside.
In order to observe a particular event, an observer must pass through
the Rindler horizon of that event.
Let us consider the Rindler horizon of the event $(t=0, \vec{x} = 0)$, i.e.,
the star at time $t=0$. (This event is simultaneous with
the event marking the beginning of the accelerating of the observer;
this is true w.r.t.\ to both the local rest systems at hand.)
If the acceleration is such that $a x_0 \geqslant 1$,
the world line of the observer is entirely outside of
(a subset of) the Rindler horizon, see Figures~\ref{rh2} and~\ref{rh3}.
In the special case $a x_0 = 1$, the observer's null asymptote coincides with
the horizon. Accordingly, the observer is unable to
observe the event $(t=0,\vec{x}= 0)$; the observer ``flees the scene''
so quickly as to leave the ``light behind him'' (i.e., the light
emitted by the star at the time of departure).
In contrast, if the acceleration is small, i.e., if $a x_0 < 1$,
the observer passes through the Rindler horizon, see Figure~\ref{rh1}.
It can be shown that the (proper) time at which the observer
intersects the Rindler horizon is intimately
connected with the time $s = s_m$, where the apex angle $\theta$ assumes
its maximum $\theta_m$. (In fact, if we considered the simpler geometry
of a luminous rod
instead of a luminous ball, the proper time at which $\mathcal{O}$ passes
through the Rindler horizon is exactly $s_m$.)

Summing up we see that several of the properties and results
derived analytically in this paper directly correspond
to geometric properties of the problem.

\section{Exercises}
\label{exerc}

In this paper we have analyzed the aberration of light in the 
case where the source is a spherical object (like a star). 
A useful problem to consider (e.g., in the context of a course 
on special relativity) is the case of a luminous rod
or the case of two point sources at a given distance,
where the observer is assumed to move in a direction perpendicular
to the rod (or the line connecting the two stars).
A uniformly accelerated observer is confronted with the
same effects as discussed in this paper, where some details
in the formulas change.

Another interesting task for students might be to consider not
a uniformly accelerated observer but a (more or less) realistic
relativistic rocket.



\begin{thebibliography}{99}

\bibitem{Gamov}
G.~Gamov, \textsl{Mr.\ Tompkins in Wonderland} (Cambridge Univ.\ Press, Cambridge, 1940).

\bibitem{Gold}
H.~Goldstein, \textsl{Classical Mechanics}, 2nd ed. (Addison-Wesley Publishing, 1980).

\bibitem{Penrose}
R.\ Penrose,
``The apparent shape of a relativistically moving sphere,''
Proc.\ Camb.\ Philos.\ Sec.\ {\bf 55}, 137--139 (1959).

\bibitem{Rindler}
W.~Rindler, \textsl{Relativity. Special, General and Cosmological}
(Oxford Univ.\ Press, 2001).

\bibitem{MKA}
T.\ M\"uller, A.\ King, and D.\ Adis,
``A trip to the end of the universe and the twin paradox,''
Electronic article: www.arxiv.org/abs/physics/0612126 (2006).


\bibitem{Wetal}
D.\ Weiskopf et.\ al,
``Visualization in the Einstein Year 2005: A Case
Study on Explanatory and Illustrative Visualization of Relativity and Astrophysics,''
VIS, p.\ 74, 16th IEEE Visualization 2005 (2005).


\bibitem{Scorgie}
G.\ C.\ Scorgie,
``Geometrical optics for space travellers,''
Eur.\ J.\ Phys.\ {\bf 10}, 7--13 (1989).

\bibitem{LD}
C.\ Lagoute and E.\ Davoust,
``The interstellar traveler,''
Am.\ J.\ Phys.\ {\bf 63} (3), 221--227 (1995).

\end{thebibliography}
\end{document}